\documentclass[nofootinbib]{revtex4}
\usepackage{graphicx}
\usepackage{latexsym}
\def\be{\begin{equation}}
\def\ee{\end{equation}}
\def\bea{\begin{eqnarray}}
\def\eea{\end{eqnarray}}

\begin{document}

\hfill USTC-ICTS-17-04

\title{On (in)stabilities of perturbations in mimetic models with higher derivatives}

\author{Yunlong Zheng$^{1,2}$}
\author{Liuyuan Shen$^{1,2}$}
\author{Yicen Mou$^{3}$}
\author{Mingzhe Li$^{3}$}
\email{limz@ustc.edu.cn}
\affiliation{$^{1}$Department of Physics, Nanjing University, Nanjing 210093, China}
\affiliation{$^{2}$Joint Center for Particle, Nuclear Physics and Cosmology, Nanjing 210093, China}
\affiliation{$^{3}$Interdisciplinary Center for Theoretical Study, University of Science and Technology of China, Hefei, Anhui 230026, China}

\begin{abstract}

Usually when applying the mimetic model to the early universe, higher derivative terms are needed to promote the mimetic field to be dynamical. However such models suffer from the ghost and/or the gradient instabilities and simple extensions cannot cure this pathology. We point out in this paper that it is possible to overcome this difficulty by considering the direct couplings of the higher derivatives of the mimetic field to the curvature of the spacetime.

\end{abstract}

\maketitle

\hskip 1.6cm PACS number(s): 98.80.-k \vskip 0.4cm

\section{Introduction}

The proposed mimetic model \cite{Chamseddine:2013kea}  was first considered as an extension of general relativity, in which the physical metric $g_{\mu\nu}$ is constructed in terms of an auxiliary metric $\tilde{g}_{\mu\nu}$ and a scalar field $\phi$, as follows
\be
g_{\mu\nu}=\tilde{g}_{\mu\nu}\tilde{g}^{\alpha\beta}\phi_{\alpha}\phi_{\beta}~,
\ee
where $\phi_{\alpha}\equiv \nabla_{\alpha}\phi$ denotes the covariant derivative of the scalar field with respect to spacetime.
In this way the conformal mode of gravity is isolated to the scalar field in a covariant way and the physical metric is invariant under the Weyl rescalings of the auxiliary metric. The resulted equations by varying the Einstein-Hilbert action, which is constructed from $g_{\mu\nu}$, with respect to the auxiliary metric and the scalar field contain the Einstein equation and an equation of motion for an extra scalar mode which can mimic the cold dark matter in the universe, so it is dubbed the mimetic dark matter \cite{Chamseddine:2013kea}.
Later, as shown in Refs. \cite{Golovnev:2013jxa,Barvinsky:2013mea}, without introducing the auxiliary metric the mimetic matter can also be considered as a new scalar component with the constraint $\phi_{\mu}\phi^{\mu}=1$. Such a constraint can be realized by a Lagrange multiplier $\lambda$ in the action, that is, the action can be written as,
\be
S=\int d^4x\sqrt{g}[\frac{1}{2}R+\lambda(\phi_{\mu}\phi^{\mu}-1)]+S_m~,
\ee
where $S_m$ is the action for other matter in the universe, and we have used the unit reduced Planck mass $M_p^2=1/(8\pi G_N)=1$ and the most negative signature for the metric. These two points about the mimetic model are equivalent, at least classically. We will take the second point in this paper, as have been done in most papers on this model.

The mimetic model was generalized in Ref. \cite{Chamseddine:2014vna} by offering phenomenologically a potential $V(\phi)$ to the mimetic field, so that the mimetic matter obtains a pressure $p=-V(\phi)$. With this generalization, the mimetic model has many applications in cosmology. With appropriate choice of the potential, the generalized mimetic model can provide inflation, bounce, dark energy, and so on. This stimulated many interests in the literature, for instances, its relation with disformal transformations \cite{Bekenstein:1992pj} were discussed in Refs. \cite{Deruelle:2014zza,Arroja:2015wpa,Domenech:2015tca}, the Hamiltonian analyses were given in Refs. \cite{Malaeb:2014vua,Langlois:2015skt}, it has been applied in various modified gravity models \cite{Chaichian:2014qba,Nojiri:2014zqa,Leon:2014yua,Astashenok:2015haa,Myrzakulov:2015qaa,Odintsov:2015cwa,Rabochaya:2015haa,Odintsov:2015wwp,Arroja:2015yvd,Cognola:2016gjy,Nojiri:2016ppu,Odintsov:2016oyz,Momeni:2014qta}, and has also been studied with extensive cosmological and astrophysical interests \cite{Saadi:2014jfa,Mirzagholi:2014ifa,Matsumoto:2015wja,Momeni:2015aea,Ramazanov:2015pha,Myrzakulov:2015kda,Astashenok:2015qzw,Chamseddine:2016uyr,Nojiri:2016vhu,Babichev:2016jzg,Chamseddine:2016ktu,Sadeghnezhad:2017hmr,Liu:2017puc}, and was recently reviewed in Ref. \cite{Sebastiani:2016ras}.

The mimetic field, even being offered a potential, is non-dynamical due to the mimetic constraint. Here non-dynamical means it is not a propagating degree of freedom. Hence, when applying it to the early universe, such as inflation and bouncing, the mimetic model fails to produce, through its quantum fluctuations, the primordial perturbations which seeded the formation of large scale structure of the universe. Due to the non-dynamical nature, its fluctuations cannot be quantized in a usual way. In order to circumvent this problem, higher derivative terms of the mimetic field are introduced in Ref. \cite{Chamseddine:2014vna} so that its action becomes,
\be\label{mimeticbox}
S=\int d^4x\sqrt{g}[\frac{1}{2}R+\lambda(\phi_{\mu}\phi^{\mu}-1)-V(\phi)+\alpha (\Box\phi)^2]~,
\ee
where $\Box\phi=\phi^{\mu}_{~\mu}$. This indeed promotes the mimetic field $\phi$ to be dynamical. Usually with higher derivative terms the model suffers from the Ostrogradski instability, for example the box term in the above action brings us two components, one is the normal mode and another is a ghost, e.g., its kinetic term has a wrong sign. But the mimetic constraint will remove one dynamical degree. The question is whether the unwanted ghost mode has been removed by the mimetic constraint in the model (\ref{mimeticbox})? It was shown in Ref. \cite{Chamseddine:2014vna} that by choosing appropriate coefficient $\alpha$, the equation for the scalar perturbation has the wanted wave-like form. It seems that this mimetic model with higher derivatives is healthy and succeeds in producing the right primordial perturbations. However,  a closer investigation in Ref. \cite{Ijjas:2016pad} using the action formalism showed that this model always suffers from the gradient instability in its scalar perturbation even though the scalar perturbation avoids to be a ghost in an appropriate parameter space. The gradient instability means the speed of sound is imaginary and the perturbation grows very fast at all scales, this in turn spoils the linear perturbation theory. The model (\ref{mimeticbox}) is equivalent to the the projectable
version of the Horava-Lifshitz gravity, and such instabilities were also pointed out in Ref. \cite{Ramazanov:2016xhp}. This pathology cannot be cured by simply generalizing the higher derivative terms $\alpha (\Box\phi)^2$ in the action (\ref{mimeticbox}) to an arbitrary function $f(\Box\phi)$ \cite{Firouzjahi:2017txv}. The ways out of the ghost and gradient instabilities must be found if we want a successful early universe model based on the mimetic matter. This is the purpose of this paper.
We will show that extending the current model by considering direct couplings of the higher derivative terms to the curvature are hopeful to circumvent the instability problem. The instabilities also exist in the later universe where the mimetic field is considered as an imperfect dark matter \cite{Mirzagholi:2014ifa}, and possible solutions containing similar couplings was pointed out in a very recent paper \cite{Hirano:2017zox}.

This paper is organized as follows: in section II, we will revisit the problem of instabilities in the mimetic models with higher derivative terms by analyzing a mimetic model which is slightly more general than (\ref{mimeticbox}), the quadratic actions for the scalar and tensor perturbations show that the gradient instability cannot be removed by simple extensions; In section III we will extend the model by introducing the direct couplings between the higher derivative terms and the curvature, and show that such type of models provide the possibility to remove both the ghost and gradient instabilities; section IV is our conclusion.

\section{Revisit on the instabilities in the mimetic model with higher derivative terms}

The quadratic action of the perturbations in the model (\ref{mimeticbox}) has been obtained in Ref. \cite{Ijjas:2016pad}. It was shown there this model cannot be free from both the ghost and gradient instabilities. Here we will generalize this mimetic model slightly,
\be\label{mimeticbox1}
S=\int d^4x\sqrt{g}[\frac{f(\phi)}{2}R+\lambda(\phi_{\mu}\phi^{\mu}-1)-V(\phi)+\alpha (\Box\phi)^2+\beta \phi_{\mu\nu}\phi^{\mu\nu}]~,
\ee
where we have considered the non-minimal coupling to the curvature and the function $f(\phi)$ only depends on the mimetic field. We also introduced another second order derivative term $\beta \phi_{\mu\nu}\phi^{\mu\nu}$, this term has been considered in the mimetic model in Ref. \cite{Capela:2014xta}.
In terms of the quadratic action, we will see whether these slightly generalizations can keep the model stable. The Arnowitt-Deser-Misner (ADM) formalism provides a convenient way to obtain the quadratic actions for the cosmological perturbations around the homogeneous and isotropic background.
In this formalism, the line element is written as
\be
ds^2=N^2 dt^2-h_{ij}(dx^i+N^i dt)(dx^j+N^j dt)~,
\ee
in terms of the lapse function $N$, the shift vector $N^i$ and the induced metric $h_{ij}$. The components for the inverse metric is given by
\be
g^{00}=\frac{1}{N^2}~,~g^{0i}=-\frac{N^i}{N^2}~,~g^{ij}=-h^{ij}+\frac{N^i N^j}{N^2}~,
\ee
where $h^{ij}$ is the inverse of the induced metric, i.e.,
\be
h^{ik}h_{kj}=\delta^i_j~.
\ee
The determinant of the four dimension metric is $g=-{\rm det}|g_{\mu\nu}|=N^2 {\rm det}|h_{ij}|\equiv N^2 h$.
We will use $h_{ij}$ to raise and lower the latin indices, such as $N_i=h_{ij}N^j$. The extrinsic curvature tensor and scalar are defined as
\be
K_{ij}=\frac{N_{i|j}+N_{j|i}-\dot h_{ij}}{2N}\equiv \frac{E_{ij}}{N}~,~K=-h^{ij}K_{ij}=\frac{-h^{ij}E_{ij}}{N}\equiv \frac{E}{N}~,
\ee
where dot means time derivative and the notation $|i$ means the covariant derivative defined by the metric $h_{ij}$, so that it is compatible with the induced metric, e.g., $h_{ij|k}=0$.
With these formulae one has the decomposition of the four dimensional curvature
\be
R=-^{(3)}R-\frac{E^2-E_{ij}E^{ij}}{N^2}+\frac{2}{N\sqrt{h}}\partial_0 (\frac{\sqrt{h}E}{N})-\frac{2}{N\sqrt{h}}\partial_i (\frac{\sqrt{h}EN^i}{N}+\sqrt{h}h^{ij}\partial_jN)~.
\ee
In addition, we also need the following useful results to calculate the higher derivative terms
\bea
\Gamma^0_{00}=\frac{\dot N+N^i\partial_iN}{N}-\frac{N^i N^j E_{ij}}{N^2}~,~\Gamma^0_{0i}=\frac{\partial_i N}{N}-\frac{N^j E_{ij}}{N^2}~,~\Gamma^0_{ij}=-\frac{E_{ij}}{N^2}~.
\eea

Considering these results in the model (\ref{mimeticbox1}), one can find that we have the constraints induced by the lapse function $N$, the shift vector $N^i$ and the Lagrange multiplier $\lambda$.
Varying the action to $\lambda$  gives the mimetic constraint $\phi_{\mu}\phi^{\mu}=1$.
We can also have the constraint equation by varying the action with respect to $N$, however this equation consists the Lagrange multiplier and is not useful for our purpose. So we will skip it in this paper.
The next step is to obtain the constraint equation induced by the shift vector $N^i$, this can be done straight forwardly, however the calculation is usually lengthy. There is a simpler method introduced by Maldacena in Ref. \cite{Maldacena:2002vr}, with it we can fix the gauge first. One can find the unitary gauge is most convenient. In this gauge the scalar field has no perturbation and the induced metric only relies on the dynamical component of the scalar and tensor perturbations,
\be
\delta\phi=0, ~h_{ij}=a^2 (e^{2\zeta I+\gamma})_{ij}~,
\ee
where $I$ is the $3\times 3$ unite matrix, $\zeta$ is the curvature (scalar) perturbation and $\gamma$ is the traceless and transverse matrix which represents the tensor perturbation, i.e., $\gamma_{ii}=\partial_i\gamma_{ij}=0$. Since the $\gamma$ matrix is traceless, it is easy to see that the determinant of $h_{ij}$ is $h=a^6 e^{6\zeta}$.
Within the unitary gauge, the mimetic constraint gives
\be
N=\dot\phi~,~ \dot N=\ddot\phi~,\partial_j N=0~,
\ee
this means that there is no perturbation to the lapse function. This will greatly simplify the rest calculations, especially one finds that the higher derivative terms of the mimetic field are identical to the extrinsic curvature,
\be
\Box\phi=K=\frac{E}{N}~,~\phi_{\mu\nu}=K_{\mu\nu}~,~\phi_{\mu\nu}\phi^{\mu\nu}=K_{ij}K^{ij}=\frac{E_{ij}E^{ij}}{N^2}~.
\ee
Since the lapse function is homogeneous we can fix it to $N=1$, this means we choose the cosmic time $t$, this in turn leads to $\dot \phi=1$ and vanishing higher time derivatives\footnote{Taking $N=a$ is equivalent to choosing the conformal time. In this case the higher time derivatives of $\phi$ are not vanishing and some of them should be considered in the rest calculations}.
With these considerations the action (\ref{mimeticbox1}) becomes
\bea\label{mimeticbox2}
S&=&\int d^4 x \sqrt{h}[ -\frac{f ^{~(3)}R}{2}+\frac{(2\alpha-f)E^2+(2\beta+f)E_{ij}E^{ij}}{2}+\frac{f}{\sqrt{h}}\partial_0 (\sqrt{h}E)-\frac{f}{\sqrt{h}}\partial_i (\sqrt{h}EN^i)-V]\nonumber\\
&=& \int d^4x \sqrt{h} [-\frac{f ^{~(3)}R}{2}+\frac{(2\alpha-f)E^2+(2\beta+f)E_{ij}E^{ij}}{2}-f_{\phi}E-V+{\rm divergences}]~,
\eea
where $f_{\phi}$ means the derivative of the coupling function $f$ with respect to the mimetic field.
Then we turn to the constraint equation induced by the shift vector. The action (\ref{mimeticbox2}) only contains the covariant derivative of $N^i$, as seen by the definitions of the extrinsic curvature tensor.
The variation $\delta S/\delta N^i=0$ gives the constraint equation
\be\label{equationshift}
(f-2\alpha) \partial_i E+(f+2\beta) E_{i~~|j}^{~j}=0~.
\ee
From this equation one should get the expression of the shift vector $N^i$ by the scale factor $a$ and the dynamical variables $\zeta$ and $\gamma_{ij}$ and then substitute it to the action (\ref{mimeticbox2}) to remove this constraint.
However, we can only solve Eq. (\ref{equationshift}) perturbatively. Following the method of Ref. \cite{Maldacena:2002vr} again, in order to obtain the quadratic action we need only to calculate $N^i$ up to the linear order. From the equation (\ref{equationshift}) one will soon find that
$N^i$ is vanishing at the background, so it is at least a first order variable. As a three dimensional vector, it is generally decomposed into the longitudinal part and the transverse one, e.g.,
\be
N^i=\partial_i B+S_i~,~{\rm with}~\partial_iS_i=0~.
\ee
Using the equation (\ref{equationshift}) again one obtains that $S_i=0$ at the linear order. In fact $S_i$ represents the vector perturbation, in the model considered here there is no source for it up to the linear order. So, $N^i$ is purely longitudinal. With these in mind, we may have the expression for $N^i$ from Eq. (\ref{equationshift})
\be\label{equationshift1}
(\alpha+\beta)\partial_i N^i=(3\alpha+\beta-f)\dot \zeta~.
\ee
One can see from this result that if $\alpha+\beta=0$, the scalar perturbation has no dynamics $\dot\zeta=0$. This reflects the fact mentioned before: without higher derivatives, $\alpha=\beta=0$, the metic field is not a dynamical component. This also happens for the combination $(\Box\phi)^2-\phi_{\mu\nu}\phi^{\mu\nu}$ in the action as in the Horndeski gravity \cite{Horndeski:1974wa}. This is expectable. In the Horndeski gravity, through elaborated combinations, the scalar field only provides one dynamical component even its action contains higher derivatives. Combining the Horndeski gravity with the mimetic idea, the mimetic constraint will eliminate the dynamics and degrade the dynamical component to a constraint. This is consistent with the result of Ref. \cite{Arroja:2015yvd}, in which the perturbation equations of the mimetic Horndeski gravity have been studied.
In general case, $\alpha+\beta\neq 0$, one can see from Eq. (\ref{equationshift1}) that the divergence of $N^i$ is proportional to the time derivative of the curvature perturbation and has no relation with its spatial derivatives. This is the origin of the gradient instability in the model considered in this section. Substitute Eq. (\ref{equationshift1}) to the action (\ref{mimeticbox2}) and using the background equation we get the quadratic actions for the cosmological perturbations of this model,
\be\label{quadratic}
S^{(2)}=\int d^4x a^3[\frac{f+2\beta}{8}\dot\gamma_{ij}\dot\gamma_{ij}-\frac{f}{8a^2}\partial_l\gamma_{ij}\partial_l\gamma_{ij}+\frac{(3\alpha+\beta-f)(2\beta+f)}{\alpha+\beta}\dot\zeta^2+\frac{f}{a^2}\partial_i\zeta\partial_i\zeta]~.
\ee
So, to have no ghost, we need $f+2\beta>0$, and $(3\alpha+\beta-f)/(\alpha+\beta)>0$. This is possible by appropriate choices of the function $f$ and the parameters $\alpha$ and $\beta$. However, if we requires the tensor perturbation being free from the gradient instability, $f$ should be positive, and this will cause the gradient instability in the scalar perturbation.
The quadratic action (\ref{quadratic}) tells us that this difficulty cannot be cured by simply adding more higher derivatives in the original action. This is because the spatial derivative terms of the scalar and tensor perturbations have nothing to do with higher derivative terms of the mimetic field, they are totally come from the three dimensional curvature $^{(3)}R$, as pointed out in Ref. \cite{Ijjas:2016pad} for the model(\ref{mimeticbox}).  This is easily seen in the unitary gauge. As we mentioned before, higher derivative terms of the mimetic field are identical to the extrinsic curvature terms which by definitions depend on the time derivatives of $\zeta$ and $\gamma_{ij}$ (through $\dot h_{ij}$ in the definitions of $E_{ij}$) and the divergence of the shift vector $N^i$.
The result (\ref{equationshift1}) showed that this divergence also has nothing to do with the spatial derivatives of the scalar and tensor perturbations. So, the $\alpha$ and $\beta$ terms cannot affect the spatial derivatives of $\zeta$ and$\gamma_{ij}$ in the quadratic action. Simple extensions, like the consideration in Ref. \cite{Firouzjahi:2017txv}, or just adding more higher derivative terms cannot change this result.

The conclusion of this section is that the slightly more general model (\ref{mimeticbox1}) is still plagued by instabilities.

\section{Possible mimetic model without instabilities}

To get a way out of the difficulty presented in model (\ref{mimeticbox1}) we must consider more complicated extensions. One possible extension is the modified gravity, for example, considering $F(R)$ gravity in model (\ref{mimeticbox1}). This is possible to circumvent the instability problem and deserves further studies. It can be expected that this modification brings a new dynamical component other than the mimetic field $\phi$ due to the nonlinear function $F(R)$. In fact the $F(R)$ gravity with mimetic condition has been studied in Ref. \cite{Nojiri:2014zqa}, where the mimetic field keeps to be non-dynamical and it is not necessary to promote it to be dynamical by introducing its higher derivative terms because the cosmic density perturbation can be generated by the dynamics of the scalar degree brought by $F(R)$.
In this paper we only consider the extensions on the $\phi$-sector.
We can see from the analysis in the previous section that it is possible to overcome the difficulty by considering more complex extension in which the curvature couples to the higher derivative terms directly. For example, we may consider the following model,
\be\label{mimetic}
S=\int d^4x \sqrt{g}[\frac{f(\phi, \Box\phi)}{2}R+\lambda (\phi_{\mu}\phi^{\mu}-1)-V(\phi)+\alpha (\Box\phi)^2+\beta \phi_{\mu\nu}\phi^{\mu\nu}]~.
\ee
Different from the model discussed in the previous section, in this model the coupling function $f$ also depends on the higher derivative terms, and here for simplicity we only consider the dependence on $\Box\phi$, extension to the dependences on other higher derivative terms is straight forward.
The gravitational field equation from this action is
\bea \label{eom1}
f G^{\mu\nu} &=& -2\lambda \phi^{\mu}\phi^{\nu}+\Box f g^{\mu\nu}-f^{\mu\nu}-Vg^{\mu\nu} -\phi^{\rho}\nabla_{\rho}(\frac{f_E}{2}R+2\alpha \Box\phi)g^{\mu\nu}-(\frac{f_E}{2}R+2\alpha \Box\phi)\Box\phi g^{\mu\nu}
-2\beta \phi^{\mu\nu}_{~~~\rho}\phi^{\rho}-2\beta \phi^{\mu\nu}\Box\phi\nonumber\\
&+&\nabla^{\mu}(\frac{f_E}{2}R+2\alpha \Box\phi)\phi^{\nu}
+\nabla^{\nu}(\frac{f_E}{2}R+2\alpha \Box\phi)\phi^{\mu}+2\beta (\Box \phi^{\mu} \phi^{\nu}+\Box \phi^{\nu}\phi^{\mu})+\alpha (\Box\phi)^2 g^{\mu\nu}+\beta \phi_{\rho\sigma}\phi^{\rho\sigma} g^{\mu\nu}~,
\eea
where the Lagrange multiplier $\lambda$ can be obtained by taking the trace on above equation,
\bea \label{eom2}
2\lambda&=&-fG +3\Box f-4V-2\phi^{\rho}\nabla_{\rho}(\frac{f_E}{2}R+2\alpha \Box\phi)-4 (\frac{f_E}{2}R+\alpha \Box\phi)\Box\phi-2\beta \phi^{\rho}\nabla_{\rho}\Box\phi-2\beta (\Box\phi)^2\nonumber\\
&+&4\beta \phi_{\mu}\Box\phi^{\mu}+4\beta \phi_{\rho\sigma}\phi^{\rho\sigma}~,
\eea
where $G$ is the trace of the Einstein tensor $G^{\mu\nu}$. The notation $f_E$ denotes the partial derivative of $f$ with respect to $\Box\phi$, because within the unitary gauge the mimetic constraint renders $\Box\phi=E$.

By the same procedure used in the previous section, the action has the form
\bea
S&=&\int d^4 x \sqrt{h}[ -\frac{f ^{~(3)}R}{2}+\frac{(2\alpha-f)E^2+(2\beta+f)E_{ij}E^{ij}}{2}+\frac{f}{\sqrt{h}}\partial_0 (\sqrt{h}E)-\frac{f}{\sqrt{h}}\partial_i (\sqrt{h}EN^i)-V]\nonumber\\
&=& \int d^4x \sqrt{h} [-\frac{f ^{~(3)}R}{2}+\frac{(2\alpha-f)E^2+(2\beta+f)E_{ij}E^{ij}}{2}+f~(\dot E-N^i E_{|i}+E^2)-V]\nonumber\\
&=& \int d^4 x \sqrt{h} [-\frac{f ^{~(3)}R}{2}+\frac{(2\alpha+f)E^2+(2\beta+f)E_{ij}E^{ij}}{2}-FE-F_{\phi}-V+{\rm divergences}]~,
\eea
here at the second step we have considered
\be
E=\frac{\partial_0 \sqrt{h}}{\sqrt{h}}-N^i_{|i}=\Box\phi~,
\ee
and at the third step, we have introduced the function $F(\phi, \Box\phi)$ so that $F_E=f$.

The variation with respect to $N^i$ gives the constraint equation
\be\label{constraint}
\partial_i[\frac{f_E }{2}(~^{(3)}R-E^2-E_{kl}E^{kl})-2\alpha E+F+f_{\phi}]+E_i^{~j}f_{|j}+(f+2\beta) E_{i~~|j}^{~j}=0~.
\ee
Again we will find that $N^i$ vanishes at the background and is purely longitudinal up to the first order
\be\label{divergence1}
S^i=0, ~\partial^2_i B=\partial_i N^i= \frac{(2f_E/a^2)\partial_j^2 \zeta-(6\alpha+2\beta-2f+15 H f_E+18 H^2 f_{EE}-3f_{\phi E})\dot \zeta}{\theta}~.
\ee
In this formula, we used the notation convention that all variables (such as $f$, $f_E$ and so on) but the perturbations $\zeta$ and $\gamma_{ij}$ represent the background ones. We will continue to use this convention in the rest of this paper and hope that this will not cause any confusion. In above equation we also considered $E=3H$ in which $H=\dot a/a$ the Hubble parameter, and defined
\be
\theta=-2\alpha-2\beta+f_{\phi E}-5Hf_E-6H^2 f_{EE}~.
\ee
We can find an obvious change brought by this extended model from Eq. (\ref{divergence1}): the divergence of $N^i$ relies on both $\partial_j^2\zeta$ and $\dot\zeta$.

With these results, we first get the zeroth order action
\be
S^{(0)}=\int d^4x e^{3\mathcal{N}}[3(3\alpha+\beta+2f) \dot\mathcal{N}^2-3F\dot \mathcal{N}-F_{\phi}-V]\equiv \int d^4x e^{3\mathcal{N}} \mathcal{L}^{(0)}~,
\ee
where we have defined e-folding number $\mathcal{N}=\ln a$ so that $\dot\mathcal{N}=H$ and $\partial E/\partial \dot\mathcal{N}=3$.
In this zeroth order action, $\mathcal{N}$ is the unique dynamical variable, its variation gives the background equation,
\be\label{background}
\mathcal{L}^{(0)}=(\frac{d}{dt}+3H)[(6\alpha+2\beta+f)H+6f_E H^2-F-f_{\phi}]~.
\ee
One can easily check that it is consistent with the equations (\ref{eom1}) and (\ref{eom2}). In terms of this equation and the constraint (\ref{constraint}), we obtain
the quadratic action after removing some surface terms,
\bea\label{result}
S^{(2)}&=&\int d^4x a^3  \{\frac{f+2\beta}{8}\dot \gamma_{ij}\dot \gamma_{ij}-\frac{f}{8a^2}\partial_l\gamma_{ij}\partial_l\gamma_{ij}\nonumber\\
&+&[\frac{2(f+2\beta)}{\theta}+3](f+2\beta)\dot\zeta^2
+\frac{1}{a^3}\frac{d}{dt}[\frac{2af_E(f+2\beta)}{\theta}]\partial_i\zeta\partial_i\zeta+\frac{f}{a^2}\partial_i\zeta\partial_i\zeta+\frac{2f^2_E}{a^4\theta} (\partial_i^2\zeta)^2\}~.
\eea
We can see from this quadratic action that there are no higher time derivative terms for $\gamma_{ij}$ and $\zeta$, so there are still three propagating degrees of freedom in this model: two tensor degrees and one scalar degree. Their propagators can be easily read from this quadratic action (\ref{result}). To prevent these degrees of freedom from being ghosts and to eliminate the gradient instabilities, the following inequalities should be satisfied,
\be\label{requirement}
f+2\beta>0~,~f>0~,~\frac{2(f+2\beta)}{\theta}+3>0~,~\frac{1}{a}\frac{d}{dt}[\frac{2af_E(f+2\beta)}{\theta}]+f<0~.
\ee
Compared with the model considered in the previous section, this model offers a possibility of removing both the ghost and gradient instabilities in the scalar perturbation.
Another new feature for this model is the appearance of the higher spatial derivative term of $\zeta$. This will modify its dispersion relation in the momentum space. This term will bring new instability at higher momentum regions if $\theta$ is positive. This is similar to the tachyonic instability. In order to get rid of it at all scales we need an extra requirement that $\theta<0$. Combined it with the requirements in Eq. (\ref{requirement}), the parameter space is constrained by
\be\label{requirement2}
f+2\beta>0~,~f>0~,~\theta<-\frac{2}{3}(f+2\beta)~,~\frac{1}{a}\frac{d}{dt}[\frac{2af_E(f+2\beta)}{\theta}]+f<0~.
\ee
A quadratic action similar to Eq. (\ref{result}) for scalar-tensor system without the mimetic constraint has been obtained in Ref. \cite{Cai:2016thi} within the framework of effective field theory, where the authors applied it to the non-singular bouncing universe and discussed the constraints on the coefficients from the absences of instabilities.

\begin{figure}
\includegraphics[scale=1.0]{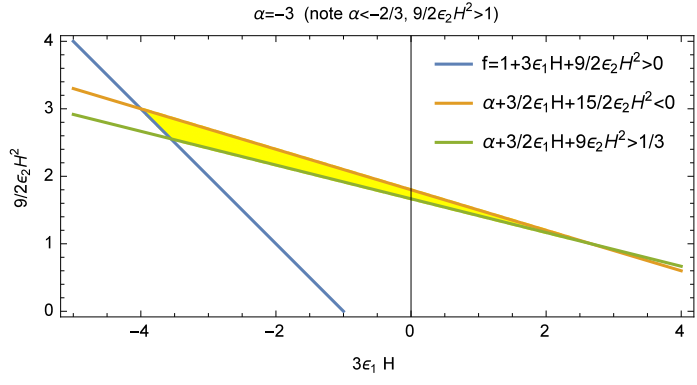}
\caption{The allowable parameter space for parameters $\epsilon_{1,2}$ if the underlying inflation model is free from all instabilities.}
\label{Fig:H}
\end{figure}

We take a toy model of inflation for an example, in which
\be
f=1+\epsilon_1\Box\phi+{1\over 2}\epsilon_2(\Box\phi)^2~,~\beta=0~,
\ee
where $\epsilon_{1,2}$ are two constant parameters and $f$ only relies on $\Box\phi$ and has no dependence on the mimetic field itself. This to some extent can be considered as a simplified version of a more realistic inflation model.
During inflation, $H$ is approximately a constant. The requirements in Eq. (\ref{requirement2}) are translated into the following inequalities,
\be
1+3\epsilon_1 H+\frac{9}{2}\epsilon_2 H^2>0~,~ \alpha+\frac{3}{2}\epsilon_1 H+9\epsilon_2 H^2-{1\over 3}>0~,~ -\alpha-\frac{3}{2}\epsilon_1 H-\frac{15}{2}\epsilon_2 H^2>0~.
\ee
One can find that the parameter $\alpha$ is constrained to be $\alpha<-2/3$, otherwise there would be no available parameter space.
After fixing $\alpha$, we can find corresponding parameter space for $\epsilon_{1,2}$, as shown by the shadow region in Fig. \ref{Fig:H}, where we take $\alpha=-3$.

In a more realistic inflation model, the spacetime of the universe during inflation is quasi de Sitter. For this we should solve the background equation
\begin{equation}
   V(\phi)=V(t)=3H^2(1-3\alpha-3\epsilon_1H-\frac{27}{2}\epsilon_{2}H^2)+2\dot{H}(1-3\alpha-\frac{9}{2}\epsilon_1H-27\epsilon_{2}H^2),
\end{equation}
from which we can easily get any background evolution including inflation  with a right potential $V(t)$.

Shift to the conformal time $\tau$ with $dt=a d\tau$,  the quadratic action (\ref{result}) is rewritten as
\begin{equation}\label{perturbation}
  S^{(2)}=\int d^3xd\tau a^{2}  \left[\frac{f}{8}(\gamma'^{2}_{ij}-\partial_l\gamma_{ij}\partial_l\gamma_{ij})+A \zeta'^{2} -B \partial_i\zeta\partial_i\zeta-C (\partial_i^2\zeta/a)^2\right]~,
\end{equation}
where the prime is the derivative with respect to $\tau$ and the coefficients in front of the scalar perturbations are defined as
\bea\label{coeffecient}
  A=(3+\frac{2f}{\theta})f~,~
  B=-\frac{2}{a^2}(\frac{aff_{E}}{\theta})'-f~,~
  C=-\frac{2f_{E}^{2}}{\theta}~,
\eea
with $\theta=-2\alpha-5\epsilon_{1}H-21\epsilon_{2}H^2~,~ f=1+3\epsilon_1H+\frac{9}{2}\epsilon_2H^2~,~f_E=\epsilon_1+3\epsilon_2H$. Again, to get rid of instabilities, all the coefficients $A, B, C$ should be positive, the sound speed square is $c_s^2=B/A$.

According to the quadratic action (\ref{perturbation}), it is straight forward to quantize the scalar and tensor perturbations and get their power spectra, as has been done in Ref. \cite{Fujita:2015ymn}, see also \cite{Martin:2002kt,Ashoorioon:2011eg,Kobayashi:2015gga,Yajima:2015xva}. Slow-roll parameter which characterize the deviation of the background from exact de Sitter background is defined to be
\be
    \epsilon_{H}=-\frac{\dot{H}}{H^2}~.
\ee
In this work we assume $\epsilon_{H}$ to be constant, which implies
\be
    a=a_*\left(\frac{\tau}{\tau_*}\right)^{-\frac{1}{1-\epsilon_H}}~,
\ee
where $a_*$ and $\tau_*$ are constants related by $a_*=-\frac{1}{(1-\epsilon_H)H_*\tau_*}$ with $H_*$ some typical Hubble scale. Other similar parameters can be defined as
\be
    \epsilon_{q}=\frac{1}{H}\frac{d\ln{q}}{dt}=\frac{d\ln{q}}{d\ln{a}}~,
\ee
which $q$ represents $c_s,A$ and other quantities. Then we have the relation
\be
    c_s(\tau)=c_{s*}\left(\frac{a}{a_*}\right)^{\epsilon_{c}}~,~A(\tau)=A_*\left(\frac{a}{a_*}\right)^{\epsilon_{A}}~.
\ee
As the above quantities are functions of Hubble parameter $H$ only, the following equation can be convenient for calculation
\be
    \epsilon_q=-\frac{d\ln{q}}{d\ln{H}}\epsilon_H~.
\ee
After some manipulation, the large scale power spectrum of the comoving curvature perturbation reads
\be
    \mathcal{P}_{\zeta}=\frac{H_*^2}{8\pi^2}\frac{\gamma}{A_*\xi c_{s*}^3}(y_*)^{(n_\zeta-1)}\left|\left(-i\gamma\xi\right)^{\mu-\nu}\frac{\Gamma(\nu)}{\Gamma(\mu)}\right|^2~,
\ee
where
\be
    y_*=-\frac{1-\epsilon_H}{1-\epsilon_H-\epsilon_C}\tau_*c_{s*}k~,~\xi=H_*\sqrt{\frac{A_*C_*}{B_*^2}}~,
\ee
and
\be
    \gamma\equiv1-\epsilon_H-\epsilon_C~,~
    \nu\equiv\frac{3-\epsilon_H+\epsilon_A}{2(1-\epsilon_H-\epsilon_C)}~,~
    \mu=\frac{\nu+1}{2}-\frac{i}{4\gamma\xi}~.
\ee
The spectral index is given by
\be
    n_\zeta-1=-\frac{2\epsilon_H+3\epsilon_C+\epsilon_A}{1-\epsilon_H-\epsilon_C}~.
\ee
Moreover, the power spectrum of the tensor perturbation is
\be
    \mathcal{P}_{t}=\frac{2H_*^2}{\pi^2f}(\tau_*k)^{n_t}~,
\ee
with
\be
    n_t=-2\epsilon_H~.
\ee
So, the tensor-to-scalar ratio is
\[r=\frac{\mathcal{P}_t}{\mathcal{P}_\zeta}\].

As an example, we numerically take the parameters $\alpha=-3,~\epsilon_1=0~,~\frac{9}{2}\epsilon_2H_*^2=1.6668$, and $\epsilon_H=0.8\times10^{-6}$ (which can be realized by choosing appropriate potential $V$), after some calculations we get the corresponding observable quantities $r=0.06$ and $n_s-1=-0.04$, these are consistent with current observational data \cite{Ade:2015lrj}. The amplitude of the scalar power spectrum depends on the value of $H_*$ which can always be achieved by choosing the potential $V(\phi)$.

\section{Conclusion}

In this paper we revisited the instability problem in the mimetic model with higher derivatives, which is important in the application of the mimetic idea to the early universe.  Using the action formalism we first studied the model which has slightly generalizations from the mimetic model considered
in Ref. \cite{Chamseddine:2014vna} and \cite{Ijjas:2016pad}. We found that this model is also plagued by the instability problem. Further naive extensions, such as adding more higher derivatives, cannot cure this pathology.
Then we pointed out that it is possible to circumvent the instability problem through introducing direct couplings of the higher derivatives of the mimetic field to the curvature of the spacetime. With such modifications the time and spatial derivatives of perturbations may have the right signs. Furthermore, these couplings will modify the dispersion relation of the scalar perturbation and in turn affect the primordial perturbations, see e.g., \cite{Cai:2009hc}. We leave the detailed discussions of the effects by the modified dispersion relation in such kind of mimetic model for future work. Other future plans include looking for the bounce universe models based on the non-minimally coupled mimetic model (\ref{mimetic}). Moreover, it would be interesting to search for more healthy models based on the direct couplings between the curvature and the higher derivatives of the mimetic matter. For instances, we may consider the following model 
\be
S\sim \int d^4x \sqrt{-{\rm det}|g_{\mu\nu}+f(\phi)\phi_{\mu\nu}|} [{1\over 2}R+\lambda(\phi_{\mu}\phi^{\mu}-1)+...]~,
\ee
where the determinant of the metric tensor is extended by including the second order derivative of the mimetic field, we may also consider the couplings of higher derivatives of the mimetic field to the $F(R)$ gravity, such as 
\be
S\sim \int d^4x \sqrt{g} [f(\phi, \Box\phi, \phi_{\mu\nu}\phi^{\mu\nu}) F(R)+\lambda(\phi_{\mu}\phi^{\mu}-1)+...]~,
\ee
this is an extension of the mimetic $F(R)$ gravity studied in Ref. \cite{Nojiri:2014zqa}. The properties and applications of these models deserve detailed investigations in the future. 

Note added: during the writing of this paper, we note that a new paper \cite{Hirano:2017zox} appeared in arXiv. In  \cite{Hirano:2017zox}, the authors considered the instability problem in the imperfect dark matter, they also pointed out from the viewpoint of effective field theory that direct couplings of the curvature to the higher derivative terms of  the mimetic field have the potential to cure the instability problem.

\section{Acknowledgement}

This work is supported in part by NSFC under Grant No. 11422543 and No. 11653002.

{}

\end{document}